\documentclass[10pt,twocolumn,twoside]{article}
\usepackage{amsfonts}
\usepackage{amssymb}
\usepackage{amsmath}
\usepackage[dvips]{graphicx}
\usepackage[latin1]{inputenc}
\usepackage{pifont}

\providecommand{\doint}{\oint}

\begin{document}

\title{The positivity and other properties of the matrix of capacitance:
physical and mathematical implications}
\author{Rodolfo A. Diaz\thanks{%
radiazs@unal.edu.co}, William J. Herrera\thanks{%
jherreraw@unal.edu.co}. \\
Departamento de Física. Universidad Nacional de Colombia. Bogotá, Colombia.}
\date{}
\maketitle

\begin{abstract}
{\small We prove that the matrix of capacitance in electrostatics is a
positive-singular matrix with a non-degenerate null eigenvalue. We explore
the physical implications of this fact, and study the physical meaning of\
the eigenvalue problem for such a matrix. Many properties are easily
visualized by constructing a \textquotedblleft potential
space\textquotedblright\ isomorphic to the euclidean space. The problem of
minimizing the internal energy of a system of conductors under constraints
is considered, and an equivalent capacitance for an arbitrary number of
conductors is obtained. Moreover, some properties of systems of conductors
in successive embedding are examined. Finally, we discuss some issues
concerning the gauge invariance of the formulation.}


\end{abstract}

\small{

\section{Introduction\label{sec:int}}

The concept of capacitance and the matrix of capacitance have been studied
from several points of view \cite{cap}-\cite{cap2}. On the other hand, the
theory of positive matrices and operators is extensively used in branches of
Physics such as the mechanics of rigid body motion, quantum mechanics \cite%
{Gold, Cohen}, and other more advanced topics \cite{topics}-\cite{topics2}.
Nevertheless, the employment of the theory of matrices and operators to
study the matrix of capacitance is rather poor \cite{Berkeley}-\cite{Grif}.
In particular, no physical meaning is usually given to the eigenvalue
problem of the matrix of capacitance. The main topic of this paper is the
proof of the fact that the matrix of capacitance is a positive matrix, as
well as the mathematical and physical consequences derived from such a fact.
The theory of positive matrices and operators permits on one hand to derive
some well-known properties of the matrix of capacitance from another point
of view, that enlighten the physical meaning of such properties. On the
other hand, it allows us to prove new mathematical properties of the matrix
of capacitance that lead to an enhancement of our theoretical understanding,
but also to new interesting applications.

The paper is distributed as follows: section \ref{sec:framework} defines the
electrostatic system of conductors that we intend to study, and establishes
the notation and properties necessary for our subsequent developments. In
Sec. \ref{sec:math prop} along with Appendix \ref{ap:positiva}, the main
goal is to prove the positivity of the matrix of capacitance. Sec. \ref%
{sec:gauge} discusses some subtleties with respect to the gauge invariance
of the formulation. Section \ref{sec:phys posit} along with appendix \ref%
{ap:uint} explores the physical implications of the positivity of the matrix
of capacitance. This is done by constructing a \textquotedblleft space of
potentials\textquotedblright\ with inner product in which the matrix of
capacitance represents an hermitian positive operator. Section \ref{sec:mini}
studies the problem of minimization of the internal energy for a system of
conductors with constraints, and an equivalent capacitance is defined for a
system with arbitrary number of conductors. On the other hand,
configurations of conductors that are successively embedded deserves special
attention because many simplifications are posible, and this is the topic of
Sec. \ref{sec:embedded} and appendix \ref{ap:embedded}. Section \ref%
{sec:conclusions} summarizes our conclusions and appendix \ref%
{ap:suggestions} contains suggested problems for readers.

\section{Basic Framework\label{sec:framework}}

This section summarizes some properties of the matrix of capacitance
obtained in Ref. \cite{AJPcapa}. They are the framework of our developments
in the remaining sections. Let us consider a system of $N$ conductors and an
equipotential surface that surrounds them, such equipotential surface could
be the cavity of an external conductor. The potential on each internal
conductor is denoted by $\varphi _{i}$, $i=1,2,\ldots ,N$. (see Fig.~\ref%
{fig:Ncond}). We define a set of surfaces $S_{i}$ slightly bigger than the
surfaces of the conductors and locally parallel to them, $\mathbf{n}_{i}$ is
an unit vector normal to the surface $S_{i}$ pointing outward with respect
to the conductor. The potential of the equipotential surface is denoted by $%
\varphi _{N+1}\ $and we define a surface $S_{N+1}\ $slightly smaller and
locally parallel to the surface of the equipotential. The charges on the
conductors are denoted by $Q_{i}$ with $i=1,...,N$ and if there is a cavity
of an external conductor in the equipotential surface we denote the charge
accumulated in such a cavity by $Q_{N+1}$, the unit vector $\mathbf{n}_{N+1}$
points inward with respect to the equipotential surface. Finally, we define
the total surface $S_{T}=S_{1}+\ldots +S_{N+1}$ and the volume $V_{S_{T}}$
defined by the surface $S_{T}$ i.e. the volume delimited by the external
surface $S_{N+1}$ and the $N$ internal surfaces $S_{i}$.

\begin{figure}[tbh]
\begin{center}
{\small \includegraphics[width=7.2cm]{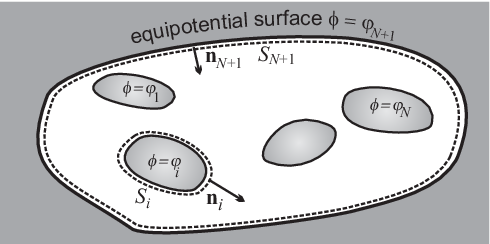} }
\end{center}
\caption{$N$ conductors surrounded by an equipotential surface. The volume $%
V_{S_{T}}$ is the region in white.}
\label{fig:Ncond}
\end{figure}

Let us define a set of dimensionless auxiliary functions $f_{i}$ that obey
Laplace's equation in the volumen $V_{S_{T}}$ with the boundary conditions 
\begin{equation}
\nabla ^{2}f_{j}=0,\text{ \ \ }f_{j}(S_{i})=\delta _{ij},\;\;(i,j=1,\ldots
,N+1).  \label{cond f}
\end{equation}%
The uniqueness theorem ensures that the solution for each $f_{j}$ is unique
in $V_{S_{T}}$. The boundary conditions (\ref{cond f}) indicate that the $%
f_{j}$ functions depend only on the geometry. Since the functions $f_{j}\ $%
acquire constant values on the surfaces $S_{i}$ with $i=1,\ldots ,N+1,\ $it
is clear that $\nabla f_{j}$ is orthogonal to these surfaces. The functions $%
f_{j}$ have some properties \cite{AJPcapa}%
\begin{equation}
\sum_{j=1}^{N+1}f_{j}=1;\ \nabla f_{j}\left( S_{i}\right) \cdot \mathbf{n}%
_{i}=\left( 1-2\delta _{ij}\right) \left\Vert \nabla f_{j}\left(
S_{i}\right) \right\Vert ;\ 0\leq f_{j}\leq 1  \label{prop f}
\end{equation}%
From these auxiliary functions we can construct a matrix that provides a
linear relation between the set of charges $Q_{i}$ and the set of potentials 
$\varphi _{i}$ in the following way%
\begin{eqnarray}
C_{ij} &\equiv &-\varepsilon _{0}\oint_{S_{i}}\nabla f_{j}\cdot \mathbf{n}%
_{i}\,dS=\varepsilon _{0}\!\int_{V_{S_{T}}}\nabla f_{i}\cdot \nabla f_{j}\,dV
\label{Cij} \\
Q_{i} &=&\sum_{j=1}^{N+1}C_{ij}\varphi _{j}  \label{QCFI}
\end{eqnarray}%
and some properties of the $C_{ij}$ matrix can be derived%
\begin{eqnarray}
C_{ij} &=&C_{ji},\ \ \sum_{j=1}^{N+1}C_{ij}=\sum_{i=1}^{N+1}C_{ij}=0,
\label{propC1} \\
C_{ii} &\geq &0\ \ ,\ \ C_{ij}\leq 0,\ \ \ \ \ \ (i\neq j).  \label{propC2}
\end{eqnarray}%
The equations above are valid for $i,j=1,\ldots ,N+1$. The expressions below
are valid for $i,j=1,\ldots ,N$%
\begin{subequations}
\label{Prop4all}
\begin{eqnarray}
\sum_{i=1}^{N}C_{i,N+1} &\leq &0\ \ ,\ \ \sum_{i=1}^{N}C_{ij}\geq 0
\label{propC3} \\
|C_{jj}| &\geq &\sum_{i\neq j}^{N}|C_{ij}|,\ \ C_{ii}C_{jj}\geq C_{ij}^{2}
\label{propC4} \\
|C_{N+1,N+1}| &=&\sum_{i=1}^{N}|C_{i,N+1}|,  \label{propC5} \\
|C_{N+1,N+1}| &\geq &|C_{i,N+1}|  \label{propC6}
\end{eqnarray}%
and expressions for the internal electrostatic energy $U\ $of the system and
of the reciprocity theorem can be obtained 
\end{subequations}
\begin{equation}
U=\frac{1}{2}\sum_{i,j}^{N+1}C_{ij}\varphi _{j}\varphi _{i}=\frac{1}{2}%
\sum_{i}^{N+1}Q_{i}\varphi _{i};\ \sum_{i=1}^{N+1}Q_{i}\varphi _{i}^{\prime
}=\sum_{j=1}^{N+1}Q_{j}^{\prime }\varphi _{j}  \label{U recip}
\end{equation}%
where $\left\{ Q_{i},\varphi _{i}\right\} $ and $\left\{ Q_{i}^{\prime
},\varphi _{i}^{\prime }\right\} $ are two sets of charges and potentials
over the same configuration of conductors. The $C_{ij}$ elements constitute
a real symmetric matrix of dimension $\left( N+1\right) \times \left(
N+1\right) $, in which the number of degrees of freedom is $N(N+1)/2$, note
that it is the same number of degrees of freedom of a $N\times N$ real
symmetric matrix.

For future purposes, we shall call the matrix with elements $C_{ij}\ $and
with $i,j=1,\ldots ,N$ the r-matrix (restricted matrix denoted by $\mathbf{C}
$), while the $C_{ij}$ matrix with $i,j=1,\ldots ,N+1$ will be called the
e-matrix (extended matrix denoted by $\mathbf{C}_{e}$).

\section{Discussion of the mathematical properties of the matrix\label%
{sec:math prop}}

In this section we establish some additional mathematical properties of the
matrix of capacitance. The central fact is that the matrix of capacitance is
a positive matrix. Basically, sections\ \ref{sec:framework} and \ref%
{sec:math prop} provide the mathematical framework whose physical
implications will be explored in the remaining sections.

Equations (\ref{cond f}) and (\ref{Cij}) tell us that the $C_{ij}$ elements
are purely geometrical. In addition, Eqs. (\ref{Cij}) and (\ref{propC1}) say
that the e-matrix is a real symmetric matrix in which the sum of elements of
each row and column is null. From Eq. (\ref{propC2}) the non-diagonal
elements of the e-matrix are non-positive. The volume integral in Eq. (\ref%
{Cij}) shows that the diagonal elements $C_{kk}$ are strictly positive for
any well-behaved geometry. In particular, since $C_{N+1,N+1}$ is positive,
Eq. (\ref{propC5}) shows that at least one element of the form $C_{i,N+1}$\
is different from zero (negative) for $i=1,\ldots ,N$; thus rewriting Eq. (%
\ref{propC1}) in the form%
\begin{equation}
\ \sum_{j=1}^{N}C_{ij}=-C_{i,N+1}  \label{anoth eq}
\end{equation}%
we see that if $C_{i,N+1}<0\ $the sum of the elements of the $i-$row of the
r-matrix is positive, if $C_{i,N+1}=0$ such a sum is null. Since at least
one of the $C_{i,N+1}$ elements is strictly negative, we conclude that in
the r-matrix the sum of elements on each row is non-negative and for at
least one row the sum is positive. Because of the symmetry, all statements
about rows are valid for columns.

On the other hand, when $V_{S_{T}}$ is a connected region as in Fig. \ref%
{fig:Ncond}, the function $f_{j}$ should change progressively from its value 
$1$ on conductor $j$ up to the value zero in the conductor $i$ without
taking local minima or maxima according to the properties of Laplace's
equation. According with Eq. (\ref{prop f}) the factor $\nabla f_{j}\left(
S_{i}\right) \cdot \mathbf{n}_{i}$ is positive for $i\neq j$ and from Eq. (%
\ref{Cij}) the non-diagonal $C_{ij}$ factors must be strictly negative for a
well-behaved geometry. This discussion is not valid when the volume $%
V_{S_{T}}$ is non-connected as in Fig. \ref{fig:embN=4}, we shall discuss
this case in section \ref{sec:embedded}. When $C_{ij}<0$ for $i\neq j$, the
discussion below Eq. (\ref{anoth eq}), leads to the fact that the sum of
elements in each row of the r-matrix is positive.

In conclusion, for the e-matrix the sum of elements of each row is null.
Further, if $V_{S_{T}}$ is a connected region, all matrix elements of the
e-matrix are non-null (for a well-behaved geometry), and for the r-matrix
the sum of elements of each row is positive. Theorems \textbf{A\ }and\ 
\textbf{B\ }in\ Appendix \ref{ap:positiva}, show that under these conditions
we find: \ding{182} The e-matrix is a real singular positive matrix, its
null eigenvalue is non-degenerate and the other eigenvalues are positive. %
\ding{183} The r-matrix is a real positive-definite matrix\footnote{%
The non-degeneration of the null eigenvalue of the e-matrix follows from
theorem \textbf{A} or alternatively from theorem \textbf{B}, in appendix \ref%
{ap:positiva}, after establishing the positive-definite nature of the
r-matrix.}. Its eigenvalues are all positive. \ding{184}\ The null
eigenvalue of the e-matrix is associated with $\left( N+1\right) -$%
dimensional eigenvectors of the form%
\begin{equation}
\mathbf{\phi }_{0}^{T}\equiv \left( \varphi _{0},\varphi _{0},...,\varphi
_{0}\right)  \label{fi const}
\end{equation}

\section{Gauge invariance of the formulation\label{sec:gauge}}

We shall see that the properties of the matrix of capacitance leads
automatically to the gauge invariance of the linear relation between charges
and potentials. An outstanding result is that the gauge invariance involving
the e-matrix, is closely related with the existence of a null eigenvalue.

We have two possible scenarios here, in the first the equipotential surface
is the surface of the cavity of a conductor that encloses the others. In the
second, the equipotential surface is just a geometrical place in the
vacuum.\ The uniqueness theorem guarantees the same solution in both cases
but only in the interior of the equipotential surface. In the equipotential
surface itself we can see that in the first case there is a charge $Q_{N+1}\ 
$accumulated in the cavity, while in the second case there is no charge in
such a surface at all. The problem lies in the fact that the electric field
is not well-behaved in the surface of the cavity because of the accumulation
of surface charge \cite{Berkeley}-\cite{Grif}, it is precisely because of
this fact that we defined surfaces slightly different from the real surfaces
on each conductor (in which $\nabla f_{j}$ are well-defined). So all the
observables (charges, potentials, electric fields) are the same in the
interior of the equipotential surface for both scenarios, but the surface
charge and the electric field differ in both cases when they are evaluated
on the equipotential surface itself\footnote{%
Of course the potential on the equipotential surface is the same in both
cases by definition.}. Anyway, the internal charges and any other
observables not defined on the equipotential surface, are calculated in both
scenarios with the same set of $C_{ij}\ $coefficients.

From the discussion above, we see that when we have a set of free
conductors, the simplest equipotential surface that we can define is the one
lying at infinity with zero potential, which is equivalent for most of the
purposes to consider a cavity of a grounded external conductor in which all
the dimensions of the cavity tend to infinity.

Further, we shall see that the linear relation between charges and
potentials in Eq. (\ref{QCFI}) is gauge invariant by shifting the potential
throughout the space as $\varphi ^{\prime }\rightarrow \varphi +\varphi _{0}$
with $\varphi _{0}$ being a non-zero constant. This gauge transformation
must keep all observables unaltered, in particular the charge $Q_{k}$ on
each surface of the conductors. Writing Eq. (\ref{QCFI}) in matrix form and
using Eq. (\ref{fi const}) we have%
\begin{equation}
\mathbf{Q}^{\prime }=\mathbf{C}_{e}\left( \mathbf{\phi +\phi }_{0}\right) =%
\mathbf{C}_{e}\mathbf{\phi =Q}  \label{gauge inv}
\end{equation}%
where we used the fact that $\mathbf{\phi }_{0}$ is an eigenvector of $%
\mathbf{C}_{e}$ with null eigenvalue. This gauge invariance says that there
is an infinite number of solutions (sets of potentials) for the linear
equations (\ref{QCFI}) with given values of the charges, this fact is
related in turn with the non-invertibility of $\mathbf{C}_{e}$. In other
words, gauge invariance is related with the existence of an eigenvector with
null eigenvalue which is also equivalent to the non-invertibility. On the
other hand, the singularity of a matrix is also related with the linear
dependence of the column (or row) vectors that constitute the matrix, this
lack of independence in the case of $\mathbf{C}_{e}$ is manifested in the
fact that no all charges can be varied independently as can be seen from the
expression%
\begin{equation}
Q_{int}=-Q_{N+1}  \label{Qind}
\end{equation}%
where $Q_{int}$ is the total charge of the internal conductors while $%
Q_{N+1} $ is the charge accumulated on the surface of the cavity of the
external conductor\footnote{%
This can be shown from Gauss's law or directly from the formalism presented
here (see Ref. \cite{AJPcapa}). If the equipotential surface is a
geometrical place in the vacuum, Eq. (\ref{Qind}) must be interpreted as a
numerical equality between the total internal charge and the quantity on the
right-hand side of Eq. (\ref{QCFI}) with $i=N+1$.}. Further, the linear
dependence of the e-matrix can be visualized by observing that it has the
same degrees of freedom as the r-matrix. This fact induces us to find
expressions involving the r-matrix only. For this, we can rewrite Eq. (\ref%
{QCFI}) by following the procedure that leads to Eq. (\ref{cijfi-fi3})%
\begin{equation}
Q_{k}=\sum_{m=1}^{N}C_{km}\left( \varphi _{m}-\varphi _{N+1}\right) \equiv
\sum_{m=1}^{N}C_{km}V_{m}  \label{QkVm}
\end{equation}%
these relations are valid for $k=1,\ldots ,N+1$. However, since Eq. (\ref%
{Qind}) shows that $Q_{N+1}$ is not independent, we can restrict them to $%
k=1,\ldots ,N$. Rewriting Eq. (\ref{QkVm}) in matrix form with this
restriction we get%
\begin{equation}
\mathbf{Q}=\mathbf{CV};\ \ \mathbf{V}\equiv \left(
V_{1},V_{2},...,V_{N}\right) ,\ \ V_{i}\equiv \varphi _{i}-\varphi _{N+1}
\label{Q=CV}
\end{equation}%
this relation is written in terms of voltages instead of potentials, so it
is clearly gauge invariant. Further, the relation is invertible because the
r-matrix $\mathbf{C}$ is positive-definite. It worths emphasizing that all
expressions obtained from now on in terms of voltages and the r-matrix, are
valid only if the voltages are taken with respect to the $\varphi _{N+1}$
potential.

\section{Physical implications of the positivity of the matrix\label%
{sec:phys posit}}

By constructing an appropriate inner product in a \textquotedblleft space of
potentials\textquotedblright , we shall derive from another point of view
some well-known results, such as the reciprocity theorem and the positive
nature of the internal energy. As a new result, we give a physical meaning
to the eigenvalues and eigenvectors of the capacitance matrix, as well as
their relation with the internal energy. Finally, we suggest some ways to
determine experimentally the set of eigenvalues and eigenvectors of $\mathbf{%
C}$, and how these eigenvectors and eigenvalues provide information about $%
\mathbf{C}$.

To facilitate the derivation and interpretation of the results let us define
the following quantities 
\begin{equation*}
c_{ij}\equiv \frac{1}{k_{0}}C_{ij}\ \ \ ;\ \ \ \Phi _{i}\equiv \frac{1}{k_{0}%
}Q_{i}
\end{equation*}%
where $k_{0}$ is a constant defined such that $c_{ij}$ are dimensionless.
From these definitions Eq. (\ref{QCFI}) could be rewritten in the form%
\begin{equation}
\Phi _{i}=\sum_{i=1}^{N+1}c_{ij}\varphi _{j}\ \ ;\ \ \mathbf{\Phi }=\mathbf{c%
}_{e}\mathbf{\phi }  \label{rot pot}
\end{equation}%
the dimensionless $c_{ij}$ factors contain the same information as $C_{ij}$.
Similarly, $\Phi _{i}$ are quantities with dimension of potential but with
the physical information of the charges $Q_{i}$ (it is like a
\textquotedblleft natural unit\textquotedblright\ for the charge). The aim
of settle the charges and potentials with the same dimension is to interpret
Eq. (\ref{rot pot}) as a linear transformation in the configuration space \ $%
\Phi ^{N+1}$ in which each axis has dimensions of potential. This space
would be isomorphic to $\mathbb{R}^{N+1}\ $if we define an inner product of
the form%
\begin{equation*}
\left( \mathbf{\Phi },\mathbf{\phi }\right) =\mathbf{\Phi }^{\dagger }%
\mathbf{\phi }=\sum_{i=1}^{N+1}\Phi _{i}\varphi _{i}
\end{equation*}%
where we have taken into account that this is a real vector space. The
capacitance matrix is hermitian (real and symmetric) with respect to this
inner product. Now let us take two sets of charges and potentials $\left\{ 
\mathbf{\Phi },\mathbf{\phi }\right\} $ and $\left\{ \mathbf{\Phi }^{\prime
},\mathbf{\phi }^{\prime }\right\} $ over the same configuration of
conductors. Doing the inner product $\left( \mathbf{\Phi }^{\prime },\mathbf{%
\phi }\right) $, using Eq. (\ref{rot pot}) and taking into account the
hermiticity of $\mathbf{c}_{e}$, we have%
\begin{equation*}
\left( \mathbf{\Phi }^{\prime },\mathbf{\phi }\right) =\left( \mathbf{c}_{e}%
\mathbf{\phi }^{\prime },\mathbf{\phi }\right) =\left( \mathbf{\phi }%
^{\prime },\mathbf{c}_{e}\mathbf{\phi }\right) =\left( \mathbf{\phi }%
^{\prime },\mathbf{\Phi }\right) =\left( \mathbf{\Phi },\mathbf{\phi }%
^{\prime }\right) 
\end{equation*}%
so that%
\begin{equation*}
\left( \mathbf{\Phi }^{\prime },\mathbf{\phi }\right) =\left( \mathbf{\Phi },%
\mathbf{\phi }^{\prime }\right) 
\end{equation*}%
which is the reciprocity theorem shown in Eq. (\ref{U recip}). From this
point of view, this theorem is a manifestation of the hermiticity of the\
e-matrix. Of course, we can define a potential space $\mathbf{\Phi }^{N}$,
in which the $N$ internal charges and $N$ voltages form $N-$dimensional
vector arrangements and the r-matrix acts as an hermitian operator. In this
space the reciprocity theorem acquires the form%
\begin{equation*}
\left( \mathbf{\Phi }^{\prime },\mathbf{V}\right) =\left( \mathbf{\Phi },%
\mathbf{V}^{\prime }\right) 
\end{equation*}%
where in this case $\mathbf{\Phi }^{\prime }$ and$\ \mathbf{\Phi }$ refer to
configurations of the internal charges only. Now we shall rewrite the
electrostatic internal energy $U\ $of the system given by Eq. (\ref{U recip}%
) in our new language%
\begin{equation}
u=\frac{1}{2}\left( \mathbf{\phi },\mathbf{c}_{e}\mathbf{\phi }\right) \geq
0\ \ \ ;\ \ u\equiv U/k_{0}  \label{uint}
\end{equation}%
the inequality comes from the positivity of the e-matrix.$\ $This expression
is gauge invariant and can be written in terms of the r-matrix and voltages
(see appendix \ref{ap:uint})\footnote{%
There is a subtlety with the concept of internal energy. The value of an
energy is not gauge invariant, but the internal energy is indeed a
difference of energies between an initial and a final configuration (or a
work to ensemble a given system) this value should then be gauge invariant.}
as follows%
\begin{equation*}
u=\frac{1}{2}\left( \mathbf{V},\mathbf{cV}\right) \geq 0
\end{equation*}%
Because $\mathbf{c}$ is positive-definite, a zero energy is obtained only
with $\mathbf{V}=\mathbf{0}$. The only configurations with zero energy are
the ones with all potentials equal\footnote{%
This is in turn related with the fact that the null eigenvalue of the
e-matrix is non-degenerate. If a degeneration of the null eigenvalue were
present, we would have at least one eigenvector associated with the zero
eigenvalue and linearly independent of the vector $\mathbf{\phi }_{0}$
defined in Eq. (\ref{fi const}). The existence of this eigenvector would
imply the existence of a configuration of different potentials with a null
value of the internal energy.}. Hence, for any geometry of the set of
conductors and for any configuration of charges and potentials on them, the
external agent that ensembles it, makes a net work on the system. There is
no configuration in which the system makes a net work on the external agent.
Note that all the analysis above is consistent with the features coming from
the equivalent equation%
\begin{equation*}
u=\frac{1}{2}\int_{V_{S_{T}}}\mathbf{E}^{2}\ dV
\end{equation*}%
where $\mathbf{E}$ is the electric field generated by the configuration
throughout the volume $V_{S_{T}}$.

Let us interpret the eigenvalue equation of $\mathbf{c}$. It reads%
\begin{equation}
\mathbf{cV}^{\left( k\right) }=\lambda _{k}\mathbf{V}^{\left( k\right) }\ \
\Rightarrow \ \ \Phi ^{\left( k\right) }=\lambda _{k}\mathbf{V}^{\left(
k\right) }  \label{charge pot prop}
\end{equation}%
we use superscripts to label a given eigenvector and subscripts to label a
given component of a fixed eigenvector. If there is a set $\left\{ i\right\}
\ $of $n$ indices such that all the $\lambda _{i}$'s are identical, this
eigenvalue is $n-$fold degenerate. According with Eq. (\ref{charge pot prop}%
), each eigenvector $\mathbf{V}^{\left( k\right) }$ means a configuration of
voltages for which each internal charge $\Phi _{i}^{\left( k\right) }\ $is
related with its corresponding voltage $V_{i}^{\left( k\right) }\ $by the
same constant of proportionality $\lambda _{k}$. Now, since the eigenvalues
are positive, each internal charge $\Phi _{i}^{\left( k\right) }$ and its
corresponding voltage $V_{i}^{\left( k\right) }$ have the same sign\footnote{%
We insist at this point that it is true only if the voltages of the internal
conductors are defined with respect to the equipotential surface that
surrounds them.}.

Let us construct a complete orthonormal set of real dimensionless
eigenvectors $\mathbf{u}^{\left( k\right) }$ of $\mathbf{c}$ associated with
the eigenvalues $\lambda _{k}$. We show in appendix \ref{ap:uint}, Eq. (\ref%
{u prop2}) that the internal energy associated with a set of voltages
described by the vector $\mathbf{V}\ $can be written in terms of those
eigenvectors and eigenvalues%
\begin{equation}
u=\frac{1}{2}\sum_{n=1}^{N}\lambda _{n}\left\vert \left( \mathbf{u}^{\left(
n\right) },\mathbf{V}\right) \right\vert ^{2}  \label{int energ val vec prop}
\end{equation}%
The set $\left\{ \mathbf{u}^{\left( n\right) }\right\} $ defines principal
axes in the potential space $\Phi ^{N}$,$\ $and $\left( \mathbf{u}^{\left(
n\right) },\mathbf{V}\right) $ is the projection of the vector $\mathbf{V}\ $%
along with the principal axis$\ \mathbf{u}^{\left( n\right) }$. If the
configuration of voltages in the system is of the form$\ \mathbf{V}^{\left(
k\right) }=V_{0}\mathbf{u}^{\left( k\right) }$ (i.e. if the vector $\mathbf{V%
}$ is parallel to a principal axis) we find\footnote{%
Since $\mathbf{u}^{\left( k\right) }$ are dimensionless, $V_{0}$ has units
of potential. Note that when $\mathbf{V\ }$is parallel to a principal axis
(i.e. becomes an eigenvector of $\mathbf{c}$), all observables become
simpler as in the case of the axis of rotation in the rigid body motion.}%
\begin{equation}
u=\frac{1}{2}\lambda _{k}V_{0}^{2}=\frac{1}{2}\lambda _{k}\left\Vert \mathbf{%
V}^{\left( k\right) }\right\Vert ^{2}  \label{int energ val vec prop2}
\end{equation}%
so the eigenvalue is proportional to the internal energy associated with a
set of voltages that forms the corresponding normalized eigenvector of the
r-matrix.

Let us suggest now a possible application that illustrates the importance of
the eigenvectors and eigenvalues of $\mathbf{c}$. Assume that for a given
configuration with $N$ internal conductors, we have calculated the matrix $%
\mathbf{c}$, as well as $N$ linearly independent eigenvectors $\mathbf{V}%
^{\left( k\right) }$ and their associated eigenvalues $\lambda _{k}$. We can
double-check the correctness of our procedure with the following experiment:
Let us settle the experimental arrangement of conductors at the voltages
defined by a given eigenvector $\mathbf{V}^{\left( p\right) }$, we then
measure the $N$ charges $\Phi _{i}^{\left( p\right) }\ $that each internal
conductor acquires. Now we calculate the $N$ quotients $\Phi _{i}^{\left(
p\right) }/V_{i}^{\left( p\right) }$ where $p$ is fixed. If our calculation
of $\mathbf{c}$ was correct, these quotients must be equal (within
experimental uncertainties) and must coincide with the eigenvalue $\lambda
_{p}$. We can proceed in the same way with each eigenvector. Further, if we
measure in each of these configurations the internal energy of the
arrangement, we can contrast these experimental values with the ones yielded
by Eq. (\ref{int energ val vec prop2}).

Though the inverse problem could be difficult in practice, it deserves to
say that we can in principle determine eigenvectors and eigenvalues
experimentally (adjusting voltages until we find constant quotients between
voltages and charges). If we can determine a complete set of eigenvectors
and eigenvalues experimentally, the matrix of capacitance can be obtained
through a similarity transformation. Defining $\mathbf{X}$ as the matrix of
eigenvectors and $\mathbf{\Lambda }$ as the matrix of eigenvalues (we use a $%
3\times 3$ matrix for illustration)

\begin{eqnarray}
\mathbf{X} &\equiv &\left( \mathbf{V}^{\left( 1\right) }\ \ \mathbf{V}%
^{\left( 2\right) }\ \ \mathbf{V}^{\left( 3\right) }\right) \equiv \left( 
\begin{array}{ccc}
V_{11} & V_{12} & V_{13} \\ 
V_{21} & V_{22} & V_{23} \\ 
V_{31} & V_{32} & V_{33}%
\end{array}%
\right)  \label{X experim} \\
\mathbf{\Lambda } &\equiv &\left( 
\begin{array}{ccc}
\lambda _{1} & 0 & 0 \\ 
0 & \lambda _{2} & 0 \\ 
0 & 0 & \lambda _{3}%
\end{array}%
\right)  \label{lamb experim}
\end{eqnarray}%
the $\mathbf{c}$ matrix can be obtained by the relation%
\begin{equation}
\mathbf{c}=\mathbf{X\Lambda X}^{-1}  \label{c experim}
\end{equation}

\setcounter{footnote}{0}

\section{Minimization of the internal energy\label{sec:mini}}

Problems of minimization of energy under constraints are very useful in
Physics. We illustrate by an example a process of minimization of the
internal electrostatic energy of a set of conductors under the contraint of
constant internal charge. The procedure followed in this section is based on
the properties of the $\mathbf{c-}$matrix developed here, and on the
Lagrange's multipliers method, leading naturally to an equivalent
capacitance between the external conductor and the set of internal
conductors. Such a procedure can be extended to more complex constraints. As
an important remark, our example shows that the e-matrix can be useful for
practical calculations, despite it does not contain additional degrees of
freedom with respect to the r-matrix.

For $N$ internal conductors inside the cavity of an external conductor, let
us find the configuration $\mathbf{V}\ $of voltages that minimizes the
internal energy with the constraint that the total internal charge $Q_{int}\ 
$is a constant $Q_{0}$. Since $Q_{int}=-Q_{N+1}$ and taking into account
that Eq. (\ref{QkVm}) is also valid for $k=N+1$, we have%
\begin{equation}
Q_{int}=-\sum_{j=1}^{N}C_{N+1,j}V_{j}=Q_{0}  \label{energ multip one1}
\end{equation}%
the function $Z\left( \mathbf{V}\right) $ that defines the constraint is%
\begin{equation}
Z\left( \mathbf{V}\right) \equiv -\sum_{j=1}^{N}C_{N+1,j}V_{j}-Q_{0}=0
\label{dZ/dn multip}
\end{equation}%
from the Lagrange's multipliers method we have%
\begin{equation}
\frac{\partial U}{\partial V_{i}}+\beta \frac{\partial Z}{\partial V_{i}}=0\
\ ;\ \ i=1,\ldots ,N  \label{ec prin multip}
\end{equation}%
where $\beta $ is the multiplier. Writing the internal energy as

\begin{equation}
U=\frac{1}{2}\left( \mathbf{V},\mathbf{CV}\right) =\frac{1}{2}%
\sum_{k=1}^{N}\sum_{j=1}^{N}V_{k}C_{kj}V_{j}  \label{u minimized}
\end{equation}%
replacing Eqs. (\ref{dZ/dn multip}, \ref{u minimized}) into Eq. (\ref{ec
prin multip}) and using the symmetry of the matrix, we find%
\begin{equation}
\sum_{j=1}^{N}C_{ij}V_{j}=\beta C_{i,N+1}\ \ ;\ \ i=1,\ldots ,N
\label{energ multip one1a}
\end{equation}%
and applying a sum over $i\ $on Eq. (\ref{energ multip one1a})%
\begin{eqnarray}
\sum_{j=1}^{N}V_{j}\sum_{i=1}^{N}C_{ij} &=&\beta \sum_{i=1}^{N}C_{i,N+1}\ ,
\label{ec prin multip1} \\
-\sum_{j=1}^{N}V_{j}C_{N+1,j} &=&-\beta C_{N+1,N+1}  \label{energ multip one}
\end{eqnarray}%
where we have used (\ref{propC1}). Subtracting Eqs. (\ref{energ multip one}, %
\ref{energ multip one1}) and solving for $\beta $ we find%
\begin{equation}
\beta =-\frac{Q_{0}}{C_{N+1,N+1}}  \label{beta multip}
\end{equation}%
Eq. (\ref{energ multip one1a}) can be rewritten as%
\begin{equation}
\mathbf{CV}=\beta \mathbf{v}_{c}\ \ ;\ \ \mathbf{v}_{c}^{T}\equiv \left(
C_{1,N+1},C_{2,N+1},...,C_{N,N+1}\right)  \label{eq cbar}
\end{equation}

For a given $\beta $, the solution of Eq. (\ref{eq cbar}) is unique because
the r-matrix $\mathbf{C}$ is invertible. It is easy to check that $\mathbf{V}%
=\left( V_{a},...,V_{a}\right) $, is a solution of Eq. (\ref{eq cbar}),
inserting this solution in Eq. (\ref{eq cbar}), we get%
\begin{eqnarray}
V_{a}\sum_{j=1}^{N}C_{ij} &=&\beta C_{i,N+1}\ ,\ i=1,\ldots ,N  \notag \\
-V_{a}C_{i,N+1} &=&\beta C_{i,N+1}\ ,\ i=1,\ldots ,N  \label{fia beta}
\end{eqnarray}%
where we have used (\ref{propC1}). From (\ref{fia beta}) we have%
\begin{equation}
V_{a}=-\beta  \label{beta volt}
\end{equation}%
Thus, the configuration of $N\ $voltages that minimizes the energy with a
fixed value of $Q_{int}$, is given by%
\begin{equation}
\mathbf{V}^{T}=\left( -\beta ,-\beta ,...,-\beta \right)  \label{pot ext min}
\end{equation}%
This kind of solution for $\mathbf{V}\ $is expected because the
configuration of minimal energy is obtained when we\ connect all the
internal conductors among them by conducting wires, this procedure clearly
keeps $Q_{int}$ constant and equates internal potentials. Since all
potentials of the interior conductors are the same, we can define a single
voltage between the external conductor and the internal ones, this voltage
is $\left\vert \beta \right\vert $.\ Since $Q_{int}=-Q_{N+1}=Q_{0}$ we can
figure out the system as equivalent to a system consisting of two conductors
with charges $\pm Q_{0}\ $and voltage $\left\vert \beta \right\vert $. Thus,
we are led naturally to an equivalent capacitance for this system of
potentials and charges%
\begin{equation*}
\left\vert Q_{0}\right\vert =C_{eq}\left\vert \beta \right\vert \ \
\Rightarrow \ \ C_{eq}=\left\vert \frac{Q_{0}}{\beta }\right\vert
=C_{N+1,N+1}
\end{equation*}%
where we have used Eq. (\ref{beta multip}). It can be checked that the
internal energy $U$ for the configuration described by (\ref{pot ext min}) is

\begin{equation}
U=\frac{1}{2}C_{N+1,N+1}V_{a}^{2}=\frac{1}{2}C_{eq}\beta ^{2}
\label{u val min}
\end{equation}%
as expected. A brief comment with respect to the e-matrix is in order. This
matrix has no additional degrees of freedom with respect to the r-matrix so
that we can formally write all results in terms of the elements $C_{ij}$ of
the r-matrix. Notwithstanding, the extended elements could be useful for
explicit calculations. Assume for instance that for the problem in the
present section, we want to calculate the total internal charge for a given
voltage of the system, and the equivalent capacitance. These calculations
can be done with the following expressions%
\begin{eqnarray*}
Q_{int}
&=&\sum_{i=1}^{N}Q_{k}=V_{a}\sum_{i=1}^{N}\sum_{j=1}^{N}C_{ij}=C_{eq}V_{a} \\
C_{eq}
&=&C_{N+1,N+1}=-\sum_{i=1}^{N}C_{i,N+1}=\sum_{i=1}^{N}\sum_{j=1}^{N}C_{ij}
\end{eqnarray*}%
Therefore, in terms of the elements of the r-matrix all the $N(N+1)/2\ $%
coefficients must be evaluated with Eq. (\ref{Cij}), to calculate $Q_{int}$
and $C_{eq}$. In contrast, by using the e-matrix, only the $C_{N+1,N+1}\ $%
coefficient should be calculated through Eq. (\ref{Cij}) to find such
observables. This difference becomes more important as $N$ increases.
Similar advantages of using the e-matrix appear in more general contexts
(see appendix A of Ref. \cite{AJPcapa}).

\section{The case of a chain of embedded conductors\label{sec:embedded}}

Let us consider a system of conductors in successive embedding as described
in Fig. \ref{fig:embN=4} and in appendix \ref{ap:embedded}. We encounter
such systems quite often in Physics. We observe that in systems as the one
in Fig. \ref{fig:Ncond}, the volume $V_{S_{T}}$ consists of a
\textquotedblleft single piece\textquotedblright , but in successively
embedded conductors such a volume consists of \textquotedblleft several
disjoint pieces\textquotedblright\ as shown in Fig. \ref{fig:embN=4}. The
main consequence coming from this difference is the fact that some elements
of the matrix of capacitance are null for embedded systems. This fact
simplifies considerably the calculation of such a matrix. However, we shall
prove that the other properties of the matrix of capacitance remains the
same.

Section \ref{sec:math prop} shows that the e-matrix is singular positive and
the r-matrix is positive-definite. The first fact was independent of the
connectivity of $V_{S_{T}}$. In contrast, the second fact was derived from
the statement that all non-diagonal $C_{ij}$ elements were strictly
negative. However, it is shown in appendix \ref{ap:embedded} that in the
case of a chain of embedded conductors (see Fig. \ref{fig:embN=4}), some of
the non-diagonal elements $C_{ij}$ are null because the volume $V_{S_{T}}$
is disconnected. Then we should check whether the r-matrix is still
positive-definite for the chain of embedded conductors.

Appendix \ref{ap:embedded} shows that elements of the form $C_{i,i\pm 1}$
are non-zero in general. Appealing to an argument analogous to the one
presented in Sec. \ref{sec:math prop} we can show that for a well-behaved
geometry of our embedded conductors, $C_{ii}>0$ and $C_{i,i\pm 1}<0$, while
the remaining elements vanish\footnote{%
Of course if $i=1\ $(or$\ i=N+1$) the element $C_{i,i-1}$ (or $C_{i,i+1}$)
does not exist. For a given $i$, at least one of them exists.}. With these
properties and the fact that the sum of elements of each row of the e-matrix
is null, we see that the sum of elements in each row of the r-matrix is null
except for the $N-$row, for which the sum is positive. Therefore, the
r-matrix of a chain of embedded conductors satisfies the conditions of
theorem \textbf{C} in appendix \ref{ap:positiva}. Consequently, for a chain
of embedded conductors, the r-matrix continues being positive-definite, and
the e-matrix is still singular positive. Combining these facts with theorem 
\textbf{B} of appendix \ref{ap:positiva}, we obtain that the null eigenvalue
of the e-matrix is non-degenerate\footnote{%
Note that in this case, theorem \textbf{A} of appendix \ref{ap:positiva}
cannot be used to establish the non-degeneration of the null eigenvalue.}.
It is again consistent with the fact that the only configuration of null
internal energy is the one associated with all conductors at the same
potential.

\section{Conclusions\label{sec:conclusions}}

We have studied an electrostatic system consisting of a set of $N$
conductors with an equipotential surface that encloses them. The associated
matrix of capacitance has dimensions $\left( N+1\right) \times \left(
N+1\right) $ (extended or e-matrix) even if the equipotential surface goes
to infinity. It is usual in the literature to work with the matrix of
dimension $N\times N$ (restricted or r-matrix), this practice is correct
only if the voltage of the conductors is taken with respect to the potential
of the equipotential surface. We prove that the e-matrix is a real positive
and singular matrix, this is consistent with the fact that gauge invariance
requires the existence of a null non-degenerate eigenvalue of this matrix.
The r-matrix is a real positive-definite matrix so all its eigenvalues are
positive.

By constructing a \textquotedblleft potential space\textquotedblright\ with
inner product, we can derive some results such as the reciprocity theorem
and the non-negativity of the electrostatic internal energy of the system,
from another point of view. A given eigenvector $\mathbf{V}^{\left( p\right)
}\ $of the r-matrix corresponds to a set of voltages $V_{i}^{\left( p\right)
}$,$\ $such that if we settle the internal conductors at these voltages, the
charges $Q_{i}^{\left( p\right) }$ acquired by each internal conductor $i$,$%
\ $are such that the quotient $\lambda _{p}=Q_{i}^{\left( p\right)
}/V_{i}^{\left( p\right) }\ $is the same for all internal conductors and
corresponds to the eigenvalue associated with $\mathbf{V}_{i}^{\left(
p\right) }$. The positivity of the eigenvalues $\lambda _{p}$ guarantees
that each charge $Q_{i}^{\left( p\right) }$ posseses the same sign as the
associated voltage $\mathbf{V}_{i}^{\left( p\right) }$. In addition, a given
eigenvalue is proportional to the internal energy associated with the set of
voltages generated by its corresponding eigenvector. Moreover, a complete
set of orthonormal eigenvectors of the r-matrix defines principal axes in
the \textquotedblleft potential space\textquotedblright . It worths
emphasizing that eigenvectors and eigenvalues can be measured
experimentally, and provide information about the matrix of capacitance.

The problem of the minimization of the internal energy is studied under the
constraint of constant value of the total internal charge. In this case we
can define an equivalent capacitance for any number of internal conductors.
From this problem, we realized that although the e-matrix has the same
degrees of freedom as the r-matrix, such extension could lead to great
simplifications of some practical calculations.

Further, systems of successive embedded conductors are analyzed showing that
some coefficients of capacitance are null for these systems, allowing an
important simplification for practical calculations. This fact is related
with connexity properties of the volume in which Laplace's equation is
considered. Moreover, we prove that for these configurations of embedded
conductors the e-matrix is still positive singular with a non-degenerate
eigenvalue and the r-matrix is positive-definite.

Finally, the properties of the matrix of capacitance shown here, can be
useful for either a formal understanding or practical calculations in
electromagnetism. It worths observing the similarity in structure between
the matrix of capacitance and the inertia tensor.

\section*{Acknowledgments}

We thank División Nacional de Investigación de Bogotá (DIB), of Universidad
Nacional de Colombia (Bogotá) for its financial support.

\appendix

\section{Some special types of matrices\label{ap:positiva}}

This appendix concerns the study of a special type of matrices. Let define $%
\Sigma _{k}$ as the sum of the elements on the $k-$row of a given matrix. We
shall make the following

\textbf{Definition}:\ A sp-matrix, is a square real matrix of finite
dimension$\ $in which $c_{mk}=c_{km}\leq 0$ for $k\neq m$, and in which $%
\Sigma _{k}$ is non-negative for all $k$. We denote with a single prime $%
\left\{ k^{\prime }\right\} $ the set of indices for which $\Sigma
_{k^{\prime }}>0$ and with double prime $\left\{ k^{\prime \prime }\right\} $
the set of indices for which$\ \Sigma _{k^{\prime \prime }}=0$. If no prime
is used, either situation could happen.

\textbf{Theorem A}:\ If $\mathbf{C}$ is a sp-matrix, then $\mathbf{C}$ is a
positive matrix with respect to the usual complex inner product. \ding{202}
If $\left\{ k^{\prime }\right\} $ is empty, the matrix is singular and
vectors of the form $\mathbf{N}_{0}^{T}=\left( n,\ldots ,n\right) $ are
eigenvectors of $\mathbf{C}$ with null eigenvalue. Further, the null
eigenvalue is non-degenerate if all elements of the matrix are non-null. %
\ding{203} If $\left\{ k^{\prime \prime }\right\} $ is empty, the matrix is
positive-definite.

\textbf{Proof}:\ We should prove that%
\begin{equation}
\mathbf{V}^{\dagger }\mathbf{CV}\geq 0  \label{bil pos}
\end{equation}%
for an arbitrary vector $\mathbf{V}$,$\ $and we should look under what
conditions exists at least one non-zero vector $\mathbf{V\ }$for which this
bilinear expression is null.\ Rewriting $\mathbf{V}=\mathbf{N}+i\mathbf{M}$
with $\mathbf{N},\ \mathbf{M}$ being real vector arrangements, and using the
symmetry of $\mathbf{C}$, the bilinear form in Eq. (\ref{bil pos})\ becomes$%
\ \mathbf{N}^{T}\mathbf{CN}+\mathbf{M}^{T}\mathbf{CM}$. Therefore, it
suffices to prove the positivity (or non-negativity) of the bilinear form
with real vector arrangements. Let $\mathbf{N}$ be a non-zero real vector,
the associated bilinear form is%
\begin{equation*}
\mathbf{N}^{T}\mathbf{CN}=\sum_{k}n_{k}c_{kk}n_{k}+\sum_{k}\sum_{m\neq
k}c_{km}~n_{k}n_{m}
\end{equation*}%
for the remaining of this appendix, we assume \textbf{that indices labeled
with different symbols are strictly different}. We rewrite the bilinear form
as

{\small 
\begin{eqnarray}
\mathbf{N}^{T}\mathbf{CN} &=&\sum_{k}\left\{ c_{kk}n_{k}^{2}\right.  \notag
\\
&&+\frac{1}{2}\sum_{m}c_{km}[n_{k}^{2}+n_{m}^{2}-(n_{k}-n_{m})^{2}]\}  \notag
\\
\mathbf{N}^{T}\mathbf{CN} &=&\sum_{k}\left\{ \left[ c_{kk}+\frac{1}{2}%
\sum_{m}c_{km}\right] n_{k}^{2}+\frac{1}{2}\sum_{m}c_{km}n_{m}^{2}\right\} 
\notag \\
&&-\frac{1}{2}\sum_{k}\sum_{m}c_{km}\left( n_{k}-n_{m}\right) ^{2}
\label{bilinear bif}
\end{eqnarray}%
}

Now, since $\Sigma _{k}=c_{kk}+\sum_{m}c_{km}\geq 0$, we have%
\begin{equation}
c_{kk}+\frac{1}{2}\sum_{m}c_{km}\geq -\frac{1}{2}\sum_{m}c_{km}
\label{sigma ineq}
\end{equation}%
it is convenient to separate the sets $\left\{ k^{\prime }\right\} $ and $%
\left\{ k^{\prime \prime }\right\} $ in Eq. (\ref{sigma ineq})%
\begin{eqnarray}
c_{k^{\prime \prime }k^{\prime \prime }}+\frac{1}{2}\sum_{m}c_{k^{\prime
\prime }m} &=&-\frac{1}{2}\sum_{m}c_{k^{\prime \prime }m}
\label{ineq partial1} \\
c_{k^{\prime }k^{\prime }}+\frac{1}{2}\sum_{m}c_{k^{\prime }m} &>&-\frac{1}{2%
}\sum_{m}c_{k^{\prime }m}  \label{ineq partial}
\end{eqnarray}%
we examine first the case in which $\left\{ k^{\prime }\right\} $ is empty.
In that case there are no equations of the type (\ref{ineq partial}), and
all indices accomplish the equation (\ref{ineq partial1}). Using (\ref{ineq
partial1}) in Eq. (\ref{bilinear bif}) and the fact that $c_{km}\leq 0$, we
find%
\begin{eqnarray*}
\mathbf{N}^{T}\mathbf{CN} &=&\sum_{k}\left\{ \left[ -\frac{1}{2}%
\sum_{m}c_{km}\right] n_{k}^{2}+\frac{1}{2}\sum_{m}c_{km}n_{m}^{2}\right\} \\
&&+\frac{1}{2}\sum_{k}\sum_{m}\left\vert c_{km}\right\vert \left(
n_{k}-n_{m}\right) ^{2}
\end{eqnarray*}%
using the symmetry of the matrix and taking into account that $k,m$ are dumb
indices, the first two terms on the right-hand side vanish and we find%
\begin{equation}
\mathbf{N}^{T}\mathbf{CN}=\frac{1}{2}\sum_{k}\sum_{m}\left\vert
c_{km}\right\vert \left( n_{k}-n_{m}\right) ^{2}\geq 0  \label{bilin1}
\end{equation}%
Equation (\ref{bilin1}) shows that the bilinear form is always non-negative
and that $\mathbf{N}^{T}\mathbf{CN}=0$ for non-zero vector arrangements of
the form $\mathbf{N}_{0}^{T}\equiv \left( n,n,\ldots ,n\right) $.
Consequently, the matrix is singular positive. We can check that $\mathbf{N}%
_{0}$ is an eigenvector of $\mathbf{C}$,$\mathbf{\ }$with null eigenvalue.
If all elements $c_{km}$ are non-null, Eq. (\ref{bilin1}) shows that this is
the only linearly independent solution, so that the zero eigenvalue is
non-degenerate.

Now we examine the case in which $\left\{ k^{\prime \prime }\right\} $ is
empty, so there are no equations of the type (\ref{ineq partial1}), and all
indices accomplish the equation (\ref{ineq partial}). Replacing (\ref{ineq
partial}) into Eq. (\ref{bilinear bif}), using the symmetry of $\mathbf{C}$,
the fact that $c_{km}\leq 0$, and that $\mathbf{N}\neq 0\ $we find%
\begin{equation}
\mathbf{N}^{T}\mathbf{CN}>\frac{1}{2}\sum_{k}\sum_{m}\left\vert
c_{km}\right\vert \left( n_{k}-n_{m}\right) ^{2}\geq 0  \label{bilin2}
\end{equation}%
and the bilinear form becomes positive if and only if $\mathbf{N}\neq 0$.
Hence, the sp-matrix is positive-definite when $\left\{ k^{\prime \prime
}\right\} $ is empty. \textbf{QED}.

\textbf{Theorem B:} Let $\mathbf{C}$ be a matrix of dimension $\left(
N+1\right) \times \left( N+1\right) $, such that $\Sigma _{i}=0$ for all
rows. This matrix has eigenvectors of the form $\mathbf{N}_{0}^{T}=\left(
n,\ldots ,n\right) $ associated with a null eigenvalue. Let $\mathbf{C}_{r}$
be the $N\times N\ $submatrix of $\mathbf{C}$ consisting of the elements $%
C_{ij}$ of\ $\mathbf{C\ }$with $i,j=1,\ldots ,N$. If $\mathbf{C}_{r}$ has no
null eigenvalues\footnote{%
If $\mathbf{C}_{r}$ is a normal matrix (or if it can be brought to the
canonical form), it is equivalent to say that $\mathbf{C}_{r}$ is
non-singular.}, the null eigenvalue of $\mathbf{C}$ is non-degenerate.

\textbf{Proof}: The condition $\Sigma _{i}=0$ for $i=1,\ldots ,N+1$ gives 
\begin{equation}
\sum_{k=1}^{N+1}C_{ik}=0\ \ ;\ \ i=1,\ldots ,N+1  \label{cond sum row}
\end{equation}%
Eigenvectors of $\mathbf{C}\ $with null eigenvalues must give%
\begin{equation}
\sum_{k=1}^{N+1}C_{ik}n_{k}=0\ \ ;\ \ i=1,...,N+1  \label{null eigensol}
\end{equation}%
assuming $n_{k}=n$ for all $k$ and using condition (\ref{cond sum row}), Eq.
(\ref{null eigensol}) is satisfied. Thus, $\mathbf{N}_{0}^{T}$ is an
eigenvector associated with a null eigenvalue. From the condition (\ref{cond
sum row}) we also find%
\begin{eqnarray}
\sum_{k=1}^{N+1}C_{ik}n_{k} &=&\sum_{k=1}^{N}C_{ik}n_{k}+C_{i,N+1}n_{N+1} 
\notag \\
&=&\sum_{k=1}^{N}C_{ik}n_{k}+\left( -\sum_{k=1}^{N}C_{ik}\right) n_{N+1} 
\notag \\
\sum_{k=1}^{N+1}C_{ik}n_{k} &=&\sum_{k=1}^{N}C_{ik}\left(
n_{k}-n_{N+1}\right) ;\ i=1,..,N+1  \label{cijfi-fi3}
\end{eqnarray}%
replacing (\ref{cijfi-fi3}) in (\ref{null eigensol}) the latter becomes%
\begin{equation}
\sum_{k=1}^{N}C_{ik}\left( n_{k}-n_{N+1}\right) =0\ ;\ \ i=1,...,N+1
\label{cijfi-fi}
\end{equation}%
in particular Eq. (\ref{cijfi-fi}) holds for $i=1,...,N$. With this
restriction Eq. (\ref{cijfi-fi}) becomes%
\begin{equation}
\mathbf{C}_{r}\mathbf{V}=0\ ;\ \ \mathbf{V}\equiv \left(
n_{1}-n_{N+1},n_{2}-n_{N+1},...,n_{N}-n_{N+1}\right)  \label{cijfi-fi2}
\end{equation}%
since $\mathbf{C}_{r}$ has no null eigenvalues, the only solution for Eq. (%
\ref{cijfi-fi2}) is $\mathbf{V}=\mathbf{0}$. Hence the only type of
solutions for $\mathbf{N}$ are of the form $\mathbf{N}=\left(
n_{N+1},n_{N+1},...,n_{N+1}\right) $ which are all linearly dependent.
Hence, the null eigenvalue is non-degenerate. It is immediate that these
solutions satisfy Eq. (\ref{cijfi-fi}) for $i=N+1$ as well. Note that $%
\mathbf{C}$ is not necessarily symmetric or real. \textbf{QED.}

\textbf{Theorem C}: Let $\mathbf{C\ }$be$\ $a$\ N\times N\ $sp-matrix such
that $\left\{ k^{\prime \prime }\right\} =\left\{ 1,2,\ldots ,N-1\right\} \ $%
and the terms%
\begin{equation*}
c_{i,i+1}=c_{i+1,i}\ \ ,\ c_{i,i-1}=c_{i-1,i}\ \ ;\ \ i=2,...,N-1
\end{equation*}%
are non-zero, while the remaining non-diagonal terms vanish. Then $\mathbf{C}
$ is positive-definite.

\textbf{Proof:} Assume $\mathbf{C}$ as singular and arrive to a
contradiction. Replacing Eqs. (\ref{ineq partial1}, \ref{ineq partial}) in
Eq. (\ref{bilinear bif}) we obtain {\small 
\begin{eqnarray*}
\mathbf{N}^{T}\mathbf{CN} &\geq &\sum_{k}\left\{ \left[ -\frac{1}{2}%
\sum_{m}c_{km}\right] n_{k}^{2}+\frac{1}{2}\sum_{m}c_{km}n_{m}^{2}\right\} \\
&&-\frac{1}{2}\sum_{k}\sum_{m}c_{km}\left( n_{k}-n_{m}\right) ^{2}
\end{eqnarray*}%
}such a replacement also shows that the equality holds if $n_{N}=0$, while
the strict inequality holds if $n_{N}\neq 0$. Using the symmetry of the
matrix and the fact that $c_{km}\leq 0$ we have%
\begin{equation*}
\mathbf{N}^{T}\mathbf{CN}\geq \frac{1}{2}\sum_{k}\sum_{m}\left\vert
c_{km}\right\vert \left( n_{k}-n_{m}\right) ^{2}
\end{equation*}%
Since $\mathbf{C}$ is singular, a non-trivial solution must exist for the
bilinear form to be null. For this, the equality must hold in this relation,
therefore $n_{N}=0$. The term on the right written in terms of the non-zero
elements of the matrix yields%
\begin{eqnarray*}
\mathbf{N}^{T}\mathbf{CN} &=&\frac{\left\vert c_{12}\right\vert }{2}\left(
n_{1}-n_{2}\right) ^{2}+\sum_{k=2}^{N-1}\frac{\left\vert
c_{k,k+1}\right\vert }{2}\left( n_{k}-n_{k+1}\right) ^{2} \\
&&+\sum_{k=2}^{N-1}\frac{\left\vert c_{k,k-1}\right\vert }{2}\left(
n_{k}-n_{k-1}\right) ^{2}+\frac{\left\vert c_{N,N-1}\right\vert }{2}%
n_{N-1}^{2}
\end{eqnarray*}%
for this expression to be zero each term in these sums must be zero. Since
all matrix elements involved in this expression are non-zero, the last sum
says that $n_{N-1}=0$, while the other sums say that $%
n_{1}=n_{2}=...=n_{N-1} $. Since $n_{N}$ was already zero, it shows that the
only solution is the trivial one, contradicting the singularity of the
matrix. \textbf{QED.}

\section{Some properties of the internal energy\label{ap:uint}}

The equation (\ref{uint}) for the internal energy $u\ $in \textquotedblleft
natural units\textquotedblright\ can be written in terms of voltages instead
of potentials with the r-matrix. By using the fact that $\mathbf{\phi }_{0}$
is an eigenvector of $\mathbf{c}_{e}$ with null eigenvalue, and the
hemiticity of $\mathbf{c}_{e}$, Eq. (\ref{uint}) becomes%
\begin{eqnarray*}
2u &=&\left( \mathbf{\phi },\mathbf{c}_{e}\left( \mathbf{\phi -\phi }%
_{0}\right) \right) =\left( \mathbf{c}_{e}\mathbf{\phi },\left( \mathbf{\phi
-\phi }_{0}\right) \right) \\
2u &=&\left( \mathbf{c}_{e}\left( \mathbf{\phi -\phi }_{0}\right) ,\left( 
\mathbf{\phi -\phi }_{0}\right) \right) =\left( \mathbf{\phi -\phi }_{0},%
\mathbf{c}_{e}\left( \mathbf{\phi -\phi }_{0}\right) \right)
\end{eqnarray*}%
defining a vector arrangement of $N+1\ $voltages $\mathbf{V}_{e}$ we find%
\begin{equation*}
2u=\left( \mathbf{V}_{e},\mathbf{c}_{e}\mathbf{V}_{e}\right) ;\ \ \mathbf{V}%
_{e}^{T}\equiv \left( V_{1},...,V_{N},V_{N+1}\right) ,\ \ V_{i}\equiv
\varphi _{i}-\varphi _{0}
\end{equation*}%
writing this bilinear form explicitly and expanding the sums we find%
\begin{eqnarray*}
u &=&\frac{1}{2}\sum_{i=1}^{N+1}\sum_{j=1}^{N+1}c_{ij}V_{i}V_{j} \\
&=&\frac{1}{2}\sum_{i=1}^{N}\sum_{j=1}^{N}c_{ij}V_{i}V_{j}+V_{N+1}K_{N+1} \\
K_{N+1} &\equiv &\sum_{i=1}^{N}c_{i,N+1}V_{i}+\frac{1}{2}c_{N+1,N+1}V_{N+1}
\end{eqnarray*}%
choosing $\varphi _{0}=\varphi _{N+1}$ we find $V_{N+1}=0$, hence%
\begin{eqnarray*}
u &=&\frac{1}{2}\sum_{i=1}^{N}\sum_{j=1}^{N}c_{ij}V_{i}V_{j}=\frac{1}{2}%
\left( \mathbf{V},\mathbf{cV}\right) \\
\mathbf{V}^{T} &\equiv &\left( V_{1},...,V_{N}\right) \ \ ,\ \ V_{i}\equiv
\varphi _{i}-\varphi _{N+1}
\end{eqnarray*}%
thus the internal energy can be written in terms of the r-matrix and the
voltages in a simple way as long as the latter are defined with respect to
the external potential $\varphi _{N+1}$. Since $\mathbf{V}$ is gauge
invariant $u$ also is.

On the other hand, remembering that we can always construct a complete
orthonormal set of real dimensionless eigenvectors $\mathbf{u}_{e}^{\left(
k\right) }$ of the e-matrix associated with the eigenvalues $\lambda
_{k}^{e} $, we can write the internal energy associated with a configuration 
$\mathbf{\phi }\ $of potentials in terms of these eigenvalues and
eigenvectors. Since the eigenvectors form a basis we can express $\mathbf{%
\phi }\ $as a linear combination of them%
\begin{equation*}
\mathbf{\phi }=\sum_{m=1}^{N+1}b_{m}\mathbf{u}_{e}^{\left( m\right) }\ \ \
;\ \ \ b_{m}\equiv \left( \mathbf{u}_{e}^{\left( m\right) },\mathbf{\phi }%
\right)
\end{equation*}%
and Eq. (\ref{uint}) becomes%
\begin{eqnarray}
2u &=&\left( \mathbf{\phi },\mathbf{c}_{e}\mathbf{\phi }\right) =\left(
\sum_{m=1}^{N+1}b_{m}\mathbf{u}_{e}^{\left( m\right) },\mathbf{c}%
_{e}\sum_{n=1}^{N+1}b_{n}\mathbf{u}_{e}^{\left( n\right) }\right)  \notag \\
&=&\sum_{m=1}^{N+1}\sum_{n=1}^{N+1}b_{m}b_{n}\left( \mathbf{u}_{e}^{\left(
m\right) },\lambda _{n}^{e}\mathbf{u}_{e}^{\left( n\right) }\right)  \notag
\\
&=&\sum_{m=1}^{N+1}\sum_{n=1}^{N+1}\lambda _{n}^{e}b_{m}b_{n}\delta
_{mn}=\sum_{n=1}^{N+1}\lambda _{n}^{e}b_{n}^{2}  \notag \\
u &=&\frac{1}{2}\sum_{n=1}^{N+1}\lambda _{n}^{e}\left\vert \left( \mathbf{u}%
^{\left( n\right) },\mathbf{\phi }\right) \right\vert ^{2}  \label{u prop}
\end{eqnarray}%
Since the eigenvectors $\mathbf{u}^{\left( k\right) }\ $and the matrix $%
\mathbf{c}_{e}\ $are dimensionless, the eigenvalues $\lambda _{k}^{e}$ also
are. It is straightforward to write this expression in terms of the r-matrix
and the voltages with respect to $\varphi _{N+1}$%
\begin{equation}
u=\frac{1}{2}\sum_{n=1}^{N}\lambda _{n}\left\vert \left( \mathbf{u}^{\left(
n\right) },\mathbf{V}\right) \right\vert ^{2}  \label{u prop2}
\end{equation}%
where $\lambda _{n},\mathbf{u}^{\left( n\right) }$ are eigenvalues and
eigenvectors of the r-matrix

\section{Some properties of chains of embedded conductors\label{ap:embedded}}

\begin{figure}[tbh]
\begin{center}
{\small \includegraphics[width=7.2cm]{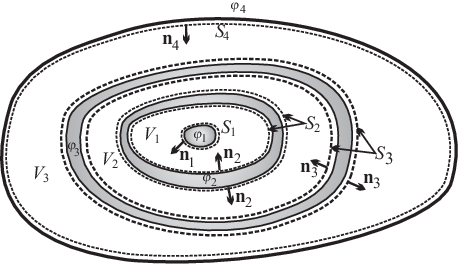} }
\end{center}
\caption{A chain of embedded conductors with $N=3$. The surfaces $S_{2}$,$%
S_{3}$, have an inner and an outer part.}
\label{fig:embN=4}
\end{figure}

Let us study a set of $N+1$ conductors which are successively embedded. We
label them $k=1,\ldots ,N+1$ from the inner to the outer. Observe that the
surface $S_{k}$ for each conductor with $k=2,\ldots ,N$ has an inner and an
outer part, but for $S_{N+1}$ we only define an inner part and for $S_{1}$
we only define an outer part (see Fig.\ \ref{fig:embN=4}). In addition, we
define $V_{k}$ with $k=1,\ldots ,N\ $as the volume formed by the points
exterior to the conductor $k$ and interior to the cavity of the conductor $%
k+1$ that contains the conductor $k$. Let us examine the non-diagonal
elements $C_{km}$ assuming from now on that $k<m$.

From Eq. (\ref{cond f}) we see that if $m-k=1$ then $f_{m}\left(
S_{k}\right) =0$ and $f_{m}\left( S_{k+1}\right) =1$ because $S_{k+1}=S_{m}$%
, the volume $V_{k}$ is precisely delimited by the outer part of the surface 
$S_{k}$ and the inner part of the surface $S_{k+1}$; thus $f_{m}$ has a
non-trivial solution in $V_{k}$. Therefore, we have in general that $\nabla
f_{m}\neq 0$ in $V_{k}$ and in the surfaces that delimite it. Thus the
integral%
\begin{equation}
C_{km}=-\varepsilon _{0}\doint\limits_{S_{k}}\nabla f_{m}\cdot \mathbf{n}%
_{k}~dS  \label{Ckm}
\end{equation}%
has a contribution from the outer part of $S_{k}$. Now, if $V_{k-1}$ exists
(i.e. if $k>1$), and taking into account that $f_{m}\left( S_{k-1}\right)
=f_{m}\left( S_{k}\right) =0$, the uniqueness theorem says that the only
solution in $V_{k-1}$ is $f_{m}=0$ and hence $\nabla f_{m}=0$ in this volume
and in the surfaces that delimite such a volume\footnote{%
Remember that the surfaces are slightly different from the surfaces of the
conductors for the gradient to be well-defined.}. Thus the integral surface
in (\ref{Ckm}) has no contributions from the inner part of $S_{k}$.

Now, if $m-k\geq 2$ we see that $f_{m}\left( S_{k}\right) =f_{m}\left(
S_{k+1}\right) =0$, then the only solution in $V_{k}$ is $f_{m}=\nabla
f_{m}=0$ in this volume and in the surfaces that delimite such a volume.
Thus the integral surface in (\ref{Ckm}) has no contributions from the inner
part of $S_{k}$. On the other hand, if $V_{k-1}$ exists ($k>1$), and since $%
f_{m}\left( S_{k-1}\right) =f_{m}\left( S_{k}\right) =0$ we see once again
that $f_{m}=\nabla f_{m}=0$ in the volume $V_{k-1}$ and in the surfaces that
delimite it; so the integral (\ref{Ckm}) has no contribution from the outer
part of $S_{k}$ either.

From the previous discussion and appealing to the symmetry of the e-matrix,
we conclude that $C_{km}=0$ for $\left\vert m-k\right\vert \geq 2$. In
addition, when $\left\vert m-k\right\vert =1$, the surface integral (\ref%
{Ckm}) receives contribution only from the outer part of $S_{k}$. Notice
that the previous behavior has to do with the fact that the total volume $%
V_{S_{T}\text{ }}$ consists of several disjoint (and so disconnected)
regions and that $\left\vert k-m\right\vert \geq 2$ indicates that these
labels are always associated with disjoint volumes. In the last discussion
we have not included the possibility that the most interior conductor has a
cavity. Since it would be an empty cavity, the surface and volume of this
cavity do not contribute to the calculation of any coefficient of
capacitance (see Ref. \cite{AJPcapa}).

From the results above, we see that for $N+1$ successively embedded
conductors with $N\geq 2$, we have%
\begin{eqnarray*}
C_{ii} &=&-\left( C_{i,i-1}+C_{i,i+1}\right) \ \ ;\ \ i=2,\ldots ,N \\
C_{11} &=&-C_{12}\ \ ;\ \ C_{N+1,N+1}=-C_{N,N+1}
\end{eqnarray*}%
How many degrees of freedom do we have for the e-matrix?.

\section{Suggested Problems\label{ap:suggestions}}

For checking the comprehension of the present formulation and its
advantages, we give some general suggestions for the reader.

\begin{enumerate}
\item Show all the properties stated here for the r-matrix and e-matrix with
specific examples.

\item Look for differences and similarities between the matrix of
capacitance in electrostatics and the inertia tensor in mechanics, from the
physical and mathematical point of view.

\item From $\left( \mathbf{\phi ,c\phi }\right) \equiv k$ we find $\left( 
\mathbf{u},\mathbf{cu}\right) =1$ with $\mathbf{u}\equiv \phi /\sqrt{k}$.
This defines the equation of an ellipsoid, describe how to find the length
of the axes of the ellipsoid in the $\Phi ^{N}$ and $\Phi ^{N+1}$ spaces for
the r-matrix and the e-matrix respectively. Describe the principal axes in
these \textquotedblleft potential spaces\textquotedblright .

\item By setting $\partial u/\partial \varphi _{i}=0$, prove that in the
absence of constraints, the only local minimum of the internal energy is
given by sets of the type $\phi _{0}$.

\item Prove Eq. (\ref{u val min}) for the minimal internal energy under the
constraint of constant internal charge.

\item Let $\left\{ a,b\right\} \ $be two positive numbers. Consider the $%
4\times 4$ matrix given by%
\begin{equation*}
\mathbf{C}_{4\times 4}=\left( 
\begin{array}{cc}
a\mathbf{B}_{2\times 2} & \mathbf{0}_{2\times 2} \\ 
\mathbf{0}_{2\times 2} & b\mathbf{B}_{2\times 2}%
\end{array}%
\right) \ \ ;\ \mathbf{B}_{2\times 2}\equiv \left( 
\begin{array}{cc}
1 & -1 \\ 
-1 & 1%
\end{array}%
\right)
\end{equation*}%
this is a sp-matrix in which the sum of elements in each row is zero.
Further, $\lambda =0$ is a two-fold degenerate eigenvalue of $\mathbf{C}$.
Can $\mathbf{C}$ be a matrix of capacitance associated with a given
electrostatic set of conductors?.

\item Look up for more applications of singular positive and
positive-definite matrices in different contexts of Physics.
\end{enumerate}

}

\end{document}